\documentclass{ws-procs9x6}

\usepackage{epsfig}

\begin{document}

\title{Highlights from ZEUS}

\author{M. Wing \footnote{\uppercase{O}n behalf of the \uppercase{ZEUS} \uppercase{C}ollaboration}}

\address{Department of Physics and Astronomy\\
University College London\\
Gower Street\\
London WC1E 6BT\\
UK\\
E-mail: mw@hep.ucl.ac.uk}

\maketitle

\abstracts{
Highlights from the ZEUS experiment at HERA as of the DIS06 conference in April are shown. 
New results focus on measurements from the 
HERA II running period of inclusive DIS at high momentum transfer 
using polarised leptons and on their impact in combined QCD and electroweak 
fits. Investigation of the hadronic final state, both for inclusive and diffractive reactions, 
continues to challenge theory. These are discussed as well as some previously 
unmeasured hadronic final states.
}

\section{Introduction}

As of April 2006, the HERA~II running period had surpassed that of the HERA I phase of 
operation: 200\,pb$^{-1}$ of data were used for measurements in this conference; 50\% 
more than in HERA~I. The ZEUS collaboration has also published its first paper using the 
HERA~II data on ``Measurement of high-$Q^2$ deep inelastic scattering cross sections 
with a longitudinally polarised positron beam at HERA''~\cite{pl:b637:210}. The results of 
this paper along with the higher-statistics measurement using incoming electrons are 
the focus of this contribution. However, the data from HERA~I, where the detector is 
well understood continues to provide a wealth of information on the 
hadronic final state, perturbative QCD and diffraction. 

\section{High $Q^2$ physics}

\begin{figure}[ht]
\centerline{\epsfysize=8cm\epsfbox{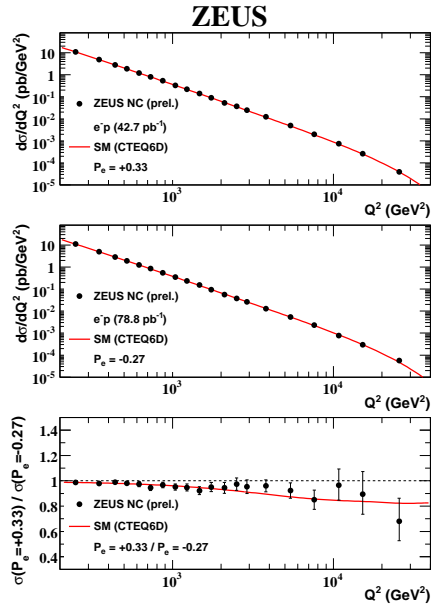}} 
\caption{The NC DIS cross section versus $Q^2$ for positive (top) and negative (middle) 
electron polarisations and the ratio of the two polarisations (bottom). The data are compared 
to the Standard Model prediction.
\label{fig:nc-cross-section}}
\end{figure}

The first HERA~II publication from ZEUS is both a measurement 
of a new process and an important step towards further results. Cross sections were measured 
in charge (CC) and neutral current (NC) DIS and are well described by Standard Model 
predictions. Although the expected strong dependence of the CC cross section on the 
polarisation of the incoming positron was observed, the far weaker dependence of the 
NC cross section was not observed at a significant level. Using a sample about 
a factor of four larger in luminosity, this effect has been observed in the recently 
taken polarised electron data~\cite{noor-dis06}, shown in Fig.~\ref{fig:nc-cross-section}.

\begin{figure}[ht]
\centerline{\epsfysize=6cm\epsfbox{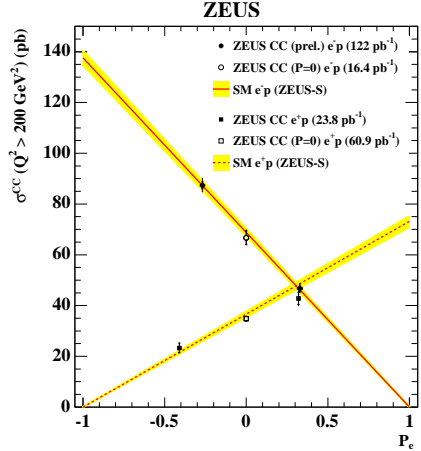}} 
\caption{Total CC DIS cross sections versus the polarisation for electrons and positrons. 
The data are compared to the Standard Model prediction. 
\label{fig:cc-cross-section}}
\end{figure}

The total CC cross section for the electron and positron data at positive and negative 
polarisations~\cite{kaji-dis06} is shown in Fig.~\ref{fig:cc-cross-section} compared 
with the unpolarised measurements from HERA~I. The data are well described by predictions 
of the Standard Model. Fitting the data and extrapolating to polarisations, $P_e = \pm 1$, 
yields values consistent with zero and hence consistent with absence of right-handed 
charge currents in the Standard Model.

Single and double differential cross sections have also been measured for both NC and 
CC processes. Using an electron data sample about eight times larger than previously, an 
improved extraction of the parity-violating structure function, $F_3$, has been made from 
the NC cross sections~\cite{noor-dis06}. However, the statistical errors on this 
measurement still dominate and so will be continually improved until the end of HERA 
running. 

All these data have been used in a combined electroweak and QCD 
fit~\cite{shimizu-dis06}, based on the previous ZEUS-JETS fit~\cite{epj:c42:1} to the 
parton densities. The new parton densities are consistent with those obtained previously 
which is a good cross check of both the new data and fit procedure. Small improvements in 
the precision of the determinations are observed for the $d$-valence, sea and gluon 
distributions. However, a significant improvement is seen for the $u$-valence as expected 
from including high-precision electron data. The QCD fit will be further improved 
with new jet measurements~\cite{jets-dis06} as well as the inclusion of other 
processes sensitive to the structure of the proton such as heavy quark and prompt photon 
production. A complete description of many different processes will provide a 
compelling test of QCD and precise measurements of the strong coupling and 
the parton density functions.

\begin{figure}[ht]
\begin{center}
~\epsfig{file=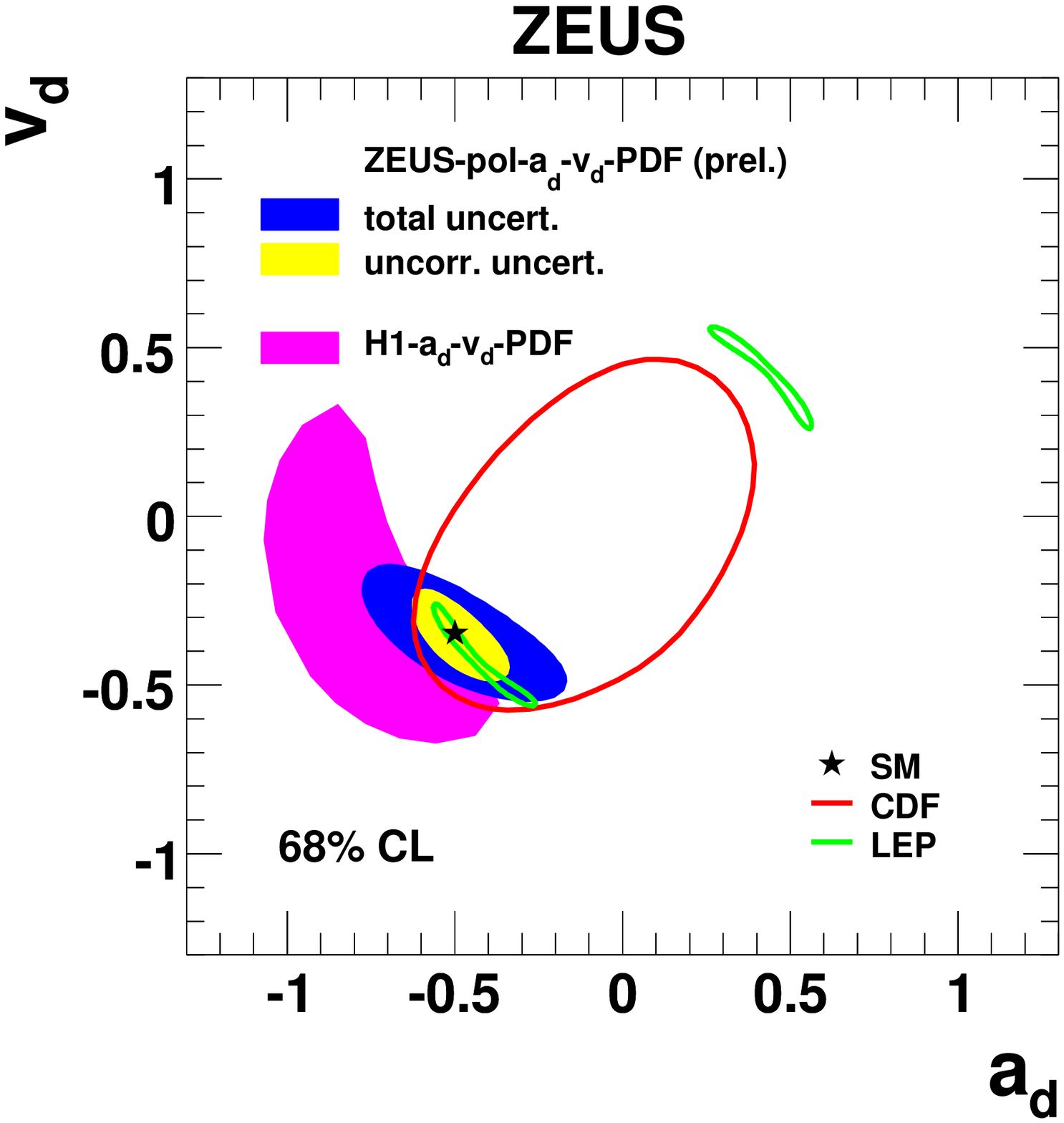,height=5cm} 
~\epsfig{file=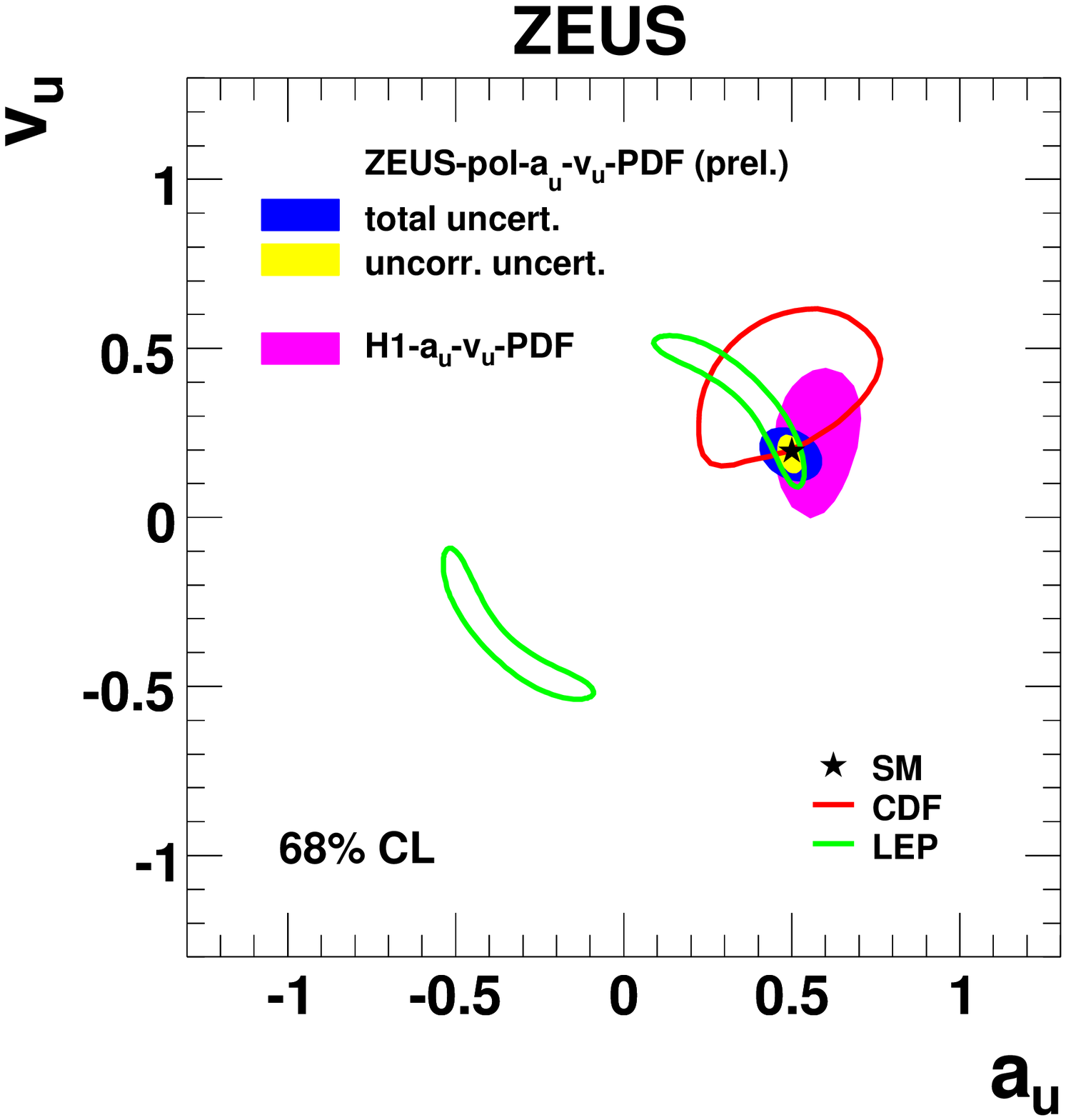,height=5cm} 
\end{center}
\caption{Axial and vector couplings of the $u$ and $d$ quarks extracted from a combined 
electroweak and QCD fit. The ZEUS measurements are compared to those from other experiments.
\label{fig:ew-fit}}
\end{figure}

Some of the electroweak parameters, the axial and vector couplings of the $u$ and $d$ 
quarks, extracted in the combined fit are shown in Fig.~\ref{fig:ew-fit}. The improved 
precision of the vector couplings due to the inclusion of polarised electron data from 
HERA~II can be seen when comparing the ZEUS result to that from H1 which only used the 
HERA~I data sample. The ZEUS measurements are also competitive, particularly for the $u$ 
quark, with those from LEP and significantly better than the CDF determinations.

As well as verifying the Standard Model over a large energy range, another aspect of 
measuring processes at the highest possible scale is to look for physics beyond the Standard 
Model. The ZEUS collaboration continues to search for events such as isolated leptons 
at high transverse momentum with a large missing transverse momentum where H1 has 
observed differences with the Standard Model prediction~\cite{perez-dis06}. ZEUS has 
adapted their analysis strategy to more closely follow that of H1 and has now analysed 
all data between 1998-2005 with an electron in the final state. In ZEUS, the rate of 
produced events is independent of the incoming lepton beam and is consistent with 
the Standard Model as can be seen in Table~\ref{tab:iso-lep}. This is somewhat in 
contrast to the results from H1 although statistical fluctuations cannot be ruled 
out.

\begin{table}[ph]
\tbl{Number of events found by H1 and ZEUS which contain isolated high $p_T$ leptons 
compared with the Standard Model and the fraction of those which are $W$ events.} 
{\footnotesize
\begin{tabular}{@{}cc@{}}
\hline
{} &{} \\[-1.5ex]
Isolated $e$ candidates &  $P_T^X > 25 $ GeV \\[1ex]
\hline
{} &{} \\[-1.5ex]
ZEUS (prel.) 98-05 $e^-p$ (143 pb$^{-1}$) & 3/2.9$\pm$0.5(53\%)  \\ [1ex]
ZEUS (prel.) 99-04 $e^+p$ (106 pb$^{-1}$) & 1/1.5$\pm$0.1(78\%)  \\ [1ex]
\hline
ZEUS (prel.) 98-05 $e^{\pm}p$ (249 pb$^{-1}$)  & 4/4.4$\pm$0.5(61\%)  \\ [1ex]
\hline
H1 (prel.)   94-05 $e^\pm p$ (279 pb$^{-1}$)   & 11/4.7 $\pm$ 0.9 (69\%)  \\ [1ex]
\hline
\end{tabular}\label{tab:iso-lep} }
\vspace*{-13pt}
\end{table}

\section{Diffraction}

\begin{figure}[ht]
\begin{center}
~\epsfig{file=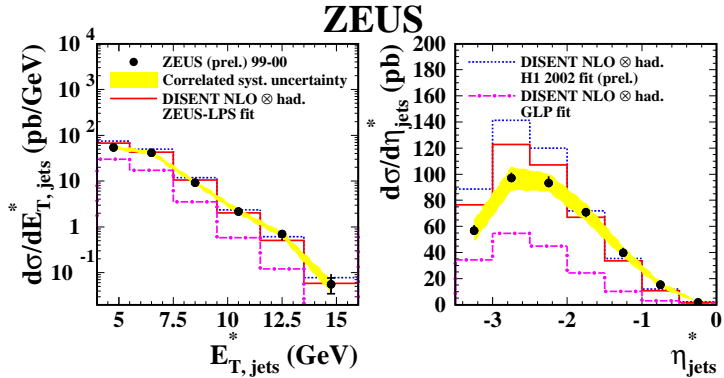,height=5cm} 
\end{center}
\caption{Cross sections for dijet production in diffractive DIS. The data are compared 
to a next-to-leading-order (NLO) QCD calculation incorporating different diffractive 
parton density functions.
\label{fig:diff-dijets}}
\end{figure}

Many recent measurements of the diffractive structure function~\cite{diff-sf} have 
allowed extractions of the diffractive parton density functions (DPDFs) to be made. 
As in inclusive production, the factorisation theorem entails 
that DPDFs extracted from one process can be used to predict the rate in another. 
This is under intensive study at HERA with measurements made of jet 
and charm rates in both photoproduction and DIS. In 
Fig.~\ref{fig:diff-dijets}, such a measurement is shown for dijet production 
in diffractive DIS compared with different DPDFs~\cite{bonato-dis06}. 
There is a 
large spread in the predictions, resulting from differences in the data to used 
make extractions of the DPDFs. This demonstrates that these dijet data are sensitive to 
the DPDFs and could be used in global fits to constrain their form. It also shows 
that more understanding of the inclusive diffractive data is needed before stronger 
statements can be made about confirmation or breaking of factorisation.

\begin{figure}[ht]
\begin{center}
~\epsfig{file=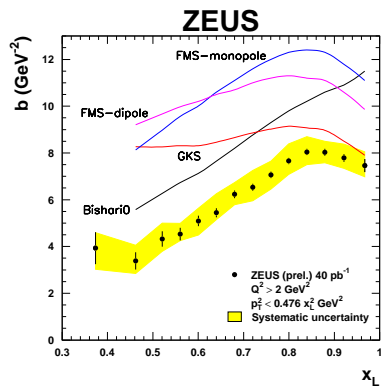,height=5cm} 
\end{center}
\caption{Measurement of the $b$ slope in leading neutron production 
compared with various models.\label{fig:ln}}
\end{figure}

Similar ideas can be probed in events with a leading neutron in the final state. If 
this process is mediated by one-pion exchange, then PDFs can also be extracted and used 
to predict rate for other processes. In the following~\cite{soares-dis06}, the production 
of leading neutrons has been measured and compared, amongst other things, to models of 
one-pion exchange. The $p_T^2$ distribution of the neutron has been measured and the 
slope of the distribution parametrised as $e^{-b p_T^2}$. The value of $b$ is plotted 
against the fraction of the beam's energy carried by the neutron, $x_L$, in 
Fig.~\ref{fig:ln}. None of the models shown (many more exist which give a poorer description 
of the data) gives a good description of the data although the general features of a rise 
to high $x_L$ and a turn-over are observed. These data have excellent discriminating power 
for these and future models.

\section{The hadronic final state}

Data from the HERA~I running period is still producing first-time and high-precision 
measurements, aided by a mature understanding of the detector.

\begin{figure}[ht]
\begin{center}
~\epsfig{file=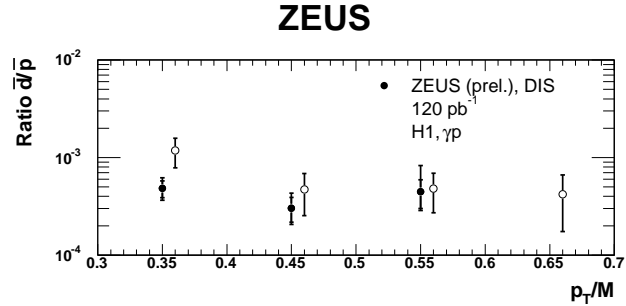,height=4cm}
\end{center}
\caption{Comparison of anti-deuteron to anti-proton production rates in 
deep inelastic scattering and photoproduction.\label{fig:deuteron}}
\end{figure}

The production light nuclei in elementary interactions is unclear and has increased 
interest recently due to the relation with pentaquark searches. Therefore a new measurement 
is presented of the observation of anti-deuteron production in DIS. The anti-deuteron 
candidate is produced at the primary interaction point and is identified using the energy 
loss measured in the central tracking detector. In Fig.~\ref{fig:deuteron}, the ratio of 
anti-deuteron and anti-proton rates are shown compared with a measurement in photoproduction 
from H1; the rates are the same in the two kinematic regions. Comparison of these data 
with e.g. the coalescence model has the potential to reveal a lot about light nuclei 
production; measurement of deuteron production should reveal more.

\begin{figure}[ht]
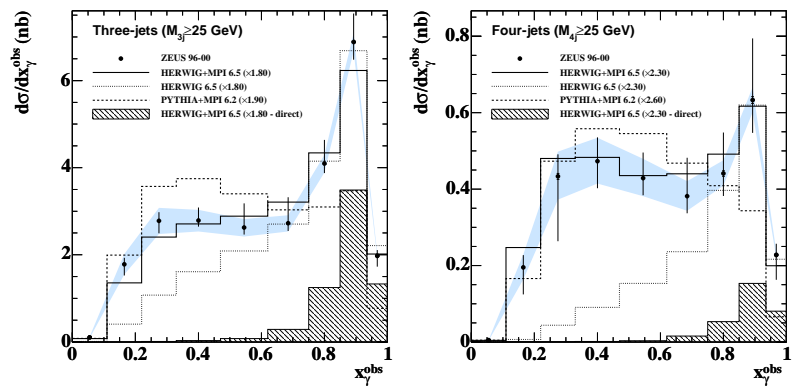

\begin{center}
~\epsfig{file=Figures/X_gamma_3j_0_HRW6505_96-00_zufo.epsi,height=5cm}
~\epsfig{file=Figures/X_gamma_4j_0_HRW6505_96-00_zufo.epsi,height=5cm} 
\end{center}
\caption{Cross section versus $x_\gamma^{\rm obs}$ for three- (left) and four-jet (right) 
photoproduction compared with predictions from Monte Carlo models.\label{fig:multijets}}
\end{figure}

Jet production in photoproduction continues to challenge QCD calculations and models. 
A recent measurement of multijet states in photoproduction~\cite{namsoo-dis06} 
intrinsically probes high-order QCD and is also sensitive to models of the underlying 
event, such as hard multi-parton interactions. Cross sections for three- and four-jet 
production are shown versus $x_\gamma^{\rm obs}$, the fraction of the photon's momentum 
participating in the multijet system, in Fig.~\ref{fig:multijets}. The cross section 
at low $x_\gamma^{\rm obs}$, where the photon is hadron-like, is significant and is 
better described by theories incorporating an underlying event model. The data have 
also been compared to a tree-level calculation - disagreements are observed - and would 
benefit enormously from a full NLO QCD calculation.

\begin{figure}[ht]
\begin{center}
~\epsfig{file=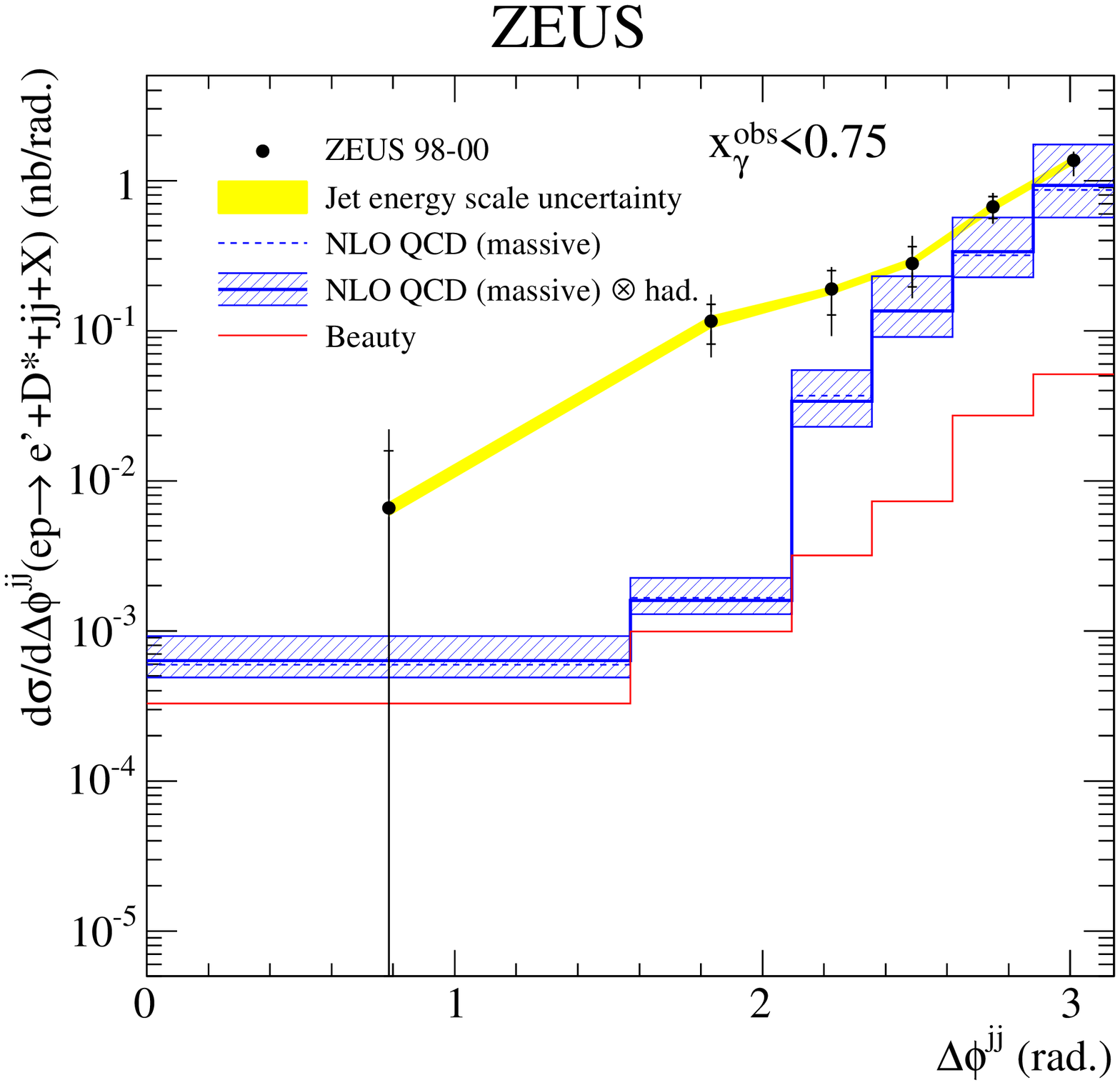,height=5cm} 
~\epsfig{file=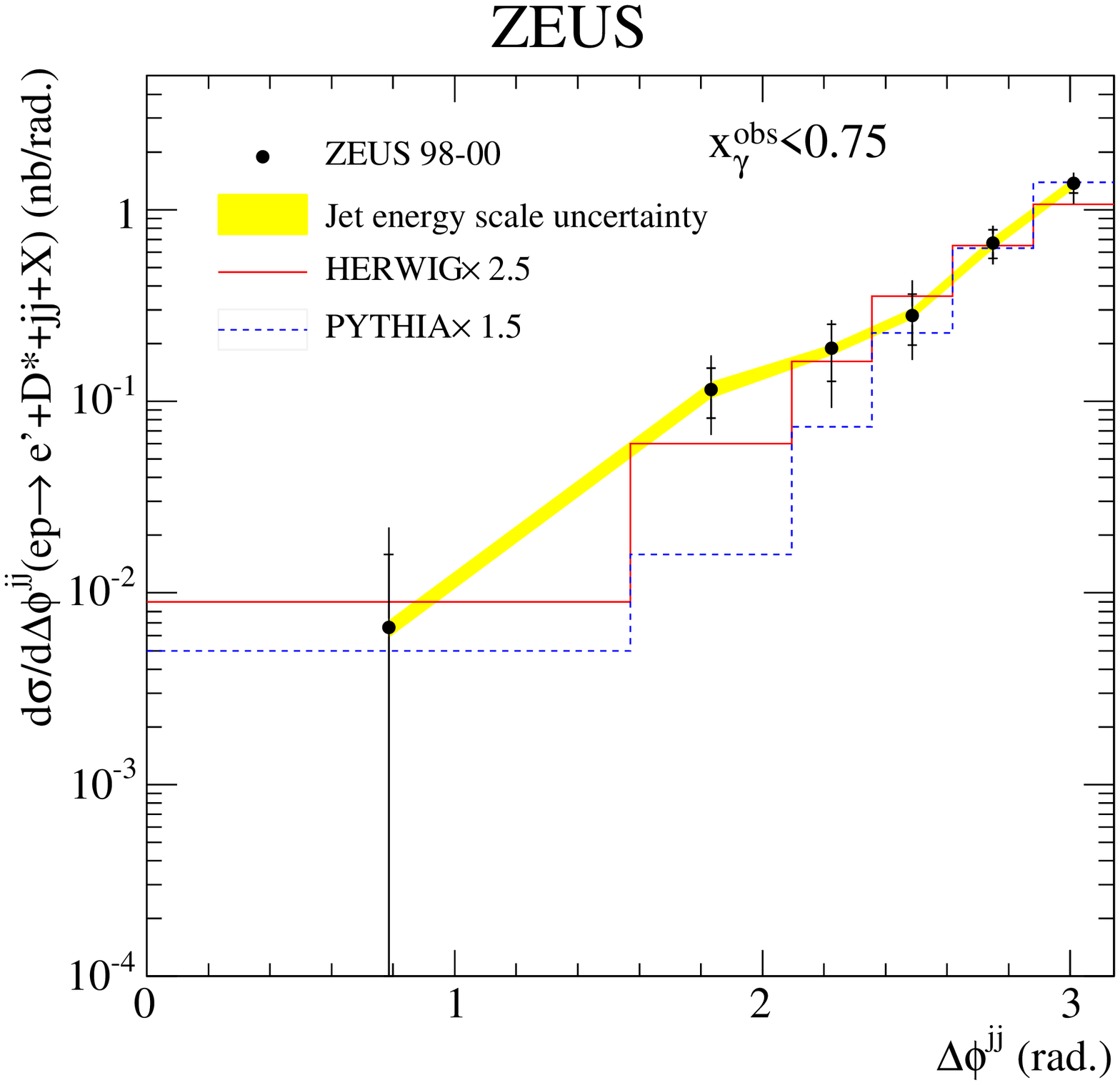,height=5cm} 
\end{center}
\caption{Charm dijet photoproduction cross sections compared to NLO QCD (left) and 
Monte Carlo models (right).
\label{fig:charm-jets}}
\end{figure}

Another area where current theoretical calculations are limited are dijet correlations 
in charm photoproduction~\cite{np:b729:492}. The inclusive-jet cross section in charm 
photoproduction 
is adequately described by NLO QCD. However, as can be seen in Fig.~\ref{fig:charm-jets}, 
the difference in azimuthal angle of two jets is poorly described for the region 
$x_\gamma^{\rm obs} < 0.75$.  The {\sc Herwig} Monte Carlo model which incorporates 
leading-order matrix elements followed by parton showers and hadronisation describes 
the shape of the data well. This indicates that for the precise description of charm 
dijet photoproduction, higher-order calculations or the implementation of additional 
parton showers in current NLO calculations are needed.

\end{document}